\documentclass{ws-mpla}
\input{babarsym}     % Standard symbols from BABAR

\newcommand{\alphas}{\alpha_s}
\newcommand{\btosg}{b\to s\gamma}
\newcommand{\BtoXclnu}{B\to\Xc\lnu}

\newcommand{\BtoXulnu}{B\to\Xu\lnu}
\newcommand{\Bptopilnu}{\Bp\to\piz\ellp\nu}
\newcommand{\Bztopilnu}{\Bz\to\pim\ellp\nu}

\newcommand{\El}{E_{\ell}}
\newcommand{\etal}{\textit{et al.}}
\newcommand{\fu}{f_u}
\newcommand{\Gthy}{\tilde{\Gamma}_{\mathrm{thy}}}
\newcommand{\Lambdabar}{\bar{\Lambda}}
\newcommand{\lambdaone}{\lambda_1}
\newcommand{\LambdaQCD}{\Lambda_{\mathrm{QCD}}}
\newcommand{\lambdatwo}{\lambda_2}
\newcommand{\lnu}{\ell\nu}
\newcommand{\mb}{m_b}
\newcommand{\mc}{m_c}

\newcommand{\muG}{\mu_G^2}
\newcommand{\mupi}{\mu_{\pi}^2}
\newcommand{\mX}{m_X}
\newcommand{\Pplus}{P_{+}}
\newcommand{\qq}{q^2}
\newcommand{\shmax}{s_h^{\max}}
\newcommand{\Xc}{X_c}
\newcommand{\Xu}{X_u}

\begin{document}

\markboth{Masahiro Morii}{Semileptonic $B$ decays and determination of \Vub}

%%%%%%%%%%%%%%%%%%%%% Publisher's Area please ignore %%%%%%%%%%%%%%%
%
\catchline{}{}{}{}{}
%
%%%%%%%%%%%%%%%%%%%%%%%%%%%%%%%%%%%%%%%%%%%%%%%%%%%%%%%%%%%%%%%%%%%%

\title{Semileptonic $B$ Decays and Determination of \Vub}

\author{\footnotesize MASAHIRO MORII}

\address{Department of Physics, Harvard University, 17 Oxford Street\\
Cambridge, Massachusetts 02138, USA\\
morii@fas.harvard.edu}

\maketitle

\pub{Received (Day Month Year)}{Revised (Day Month Year)}

\begin{abstract}
  Semileptonic decays of the $B$ mesons provide an excellent 
  probe for the weak and strong interactions of the bottom quark.
  The large data samples collected at the $B$ Factories have 
  pushed the experimental studies of the semileptonic $B$ decays
  to a new height and stimulated significant theoretical developments.
  I review recent progresses in this fast-evolving field,
  with an emphasis on the determination of the magnitude of the
  Cabibbo-Kobayashi-Maskawa matrix element \Vub.

  \keywords{Semileptonic $B$ decays; CKM matrix; \Vub}
\end{abstract}

\ccode{PACS Nos.: 12.15.Hh, 13.20.He, 14.40.Nd}

\section{Introduction}

\begin{figure}[tb]
  \begin{center}
    \psfig{file=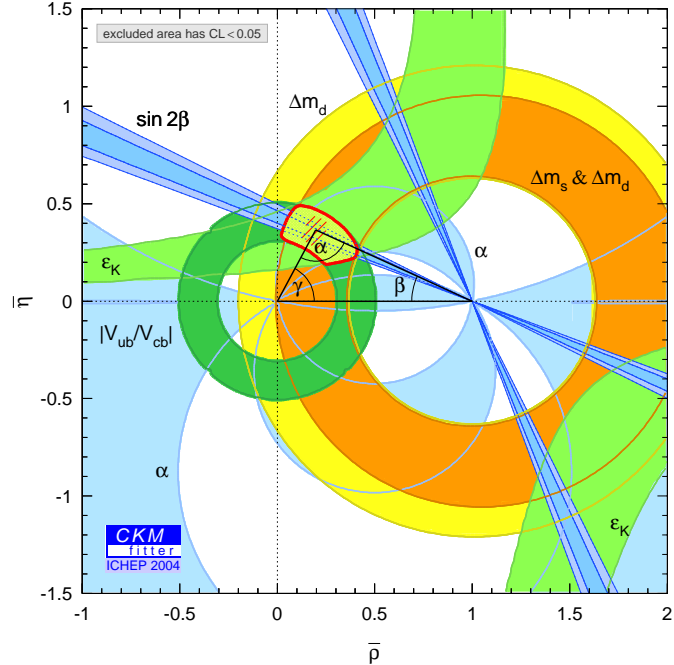,width=0.7\textwidth}
  \end{center}
  \caption[]{Global fit of the experimental data as of summer 2004
    by the CKMfitter.\cite{CKMfitter}
    The red contour contains the 95\% C.L. region based on the
    measurements of $\epsK$, $|V_{ub}/V_{cb}|$, $\Delta m_d$,
    and the lower limit on $\Delta m_s$.
    The green ring centered around $(0,0)$ is the 95\% C.L. region based on $|V_{ub}/V_{cb}|$.}
  \label{fig:ckmfitter}
\end{figure}

The success of the $B$ Factories, PEP-II and KEKB,
has dramatically improved our understanding of the
$CP$ violation.
The latest results of the time-dependent $CP$ asymmetries
in the neutral $B$ decays are in good agreement
with the predictions of the Cabibbo-Kobayashi-Maskawa mechanism\cite{KM}
as shown in Fig.~\ref{fig:ckmfitter}.
The precision of the consistency test is no longer limited by
the measurement of the $CP$-violating parameter $\sin2\beta$,
but by the measured ratio of the Cabibbo-Kobayashi-Maskawa (CKM)
matrix elements $|V_{ub}/V_{cb}|$, which determines the
length of the left side of the Unitarity Triangle.
The current uncertainties of \Vub\ and \Vcb\ are $\ge$10\%
and $\sim$2\%, respectively.
Improvement of our knowledge of \Vub\ will directly
translate to a more stringent test of the Standard Model.

Semileptonic decays of $B$ mesons provide a clean
environment for studying the tree decay amplitudes,
which allows us to determine
\Vub\ and \Vcb.
Experimental studies of charmless semileptonic $B$ decays
can be broadly categorized into \emph{inclusive} and
\emph{exclusive} measurements.
The former measures the decay rate $\Gamma(\BtoXulnu)$,
where $\Xu$ is a hadronic system without charm content.\footnote{%
  Throughout this article, the symbol $\lnu$ stands for $e^-\bar{\nu}_e$,
  $\mu^-\bar{\nu}_{\mu}$, or their charge-conjugation partners.}
The latter measures the decay rates for exclusive
final states such as $B\to\pi\lnu$ and $\rho\lnu$.
In addition to having different efficiencies and
signal-to-background ratios, the inclusive and exclusive
measurements depend on different types of theoretical
calculations.
Pursuing both approaches and comparing the results will
help us verify the robustness of the theoretical errors,
which limit the current precision of \Vub.

The experimental and theoretical issues surrounding the
determination of \Vub\ are complex and sometimes controversial.
Progress in the last few years nonetheless has made it a
concrete possibility that \Vub\ will soon be determined with
a precision of 10\%\ or better.
In this article, I review the current status of the measurements
and discuss potential improvements that can be achieved
in the near future.

\section{Inclusive Measurement of \Vub}\label{sec:incl}

In an inclusive measurement of \Vub, 
one measures the rate of the charmless semileptonic decay
$\BtoXulnu$ without reconstructing the hadronic system $\Xu$.
Since the $u$ quark is much lighter than the $c$ quark,
the $\BtoXulnu$ signal can be statistically separated
from the more copious $\BtoXclnu$ background
taking advantage of the differences in the decay kinematics.

\subsection{Theoretical Background}\label{ssec:incltheory}

Quark-hadron duality\cite{BigiDuality}
connects the inclusive decay width
$\Gamma(B\to\Xu\lnu)$ to the quark-level decay width
\begin{equation}
\Gamma(b\to u\lnu) = \frac{G_F^2\Vub^2}{192\pi^3}\mb^5.
\label{eq:Gquark}
\end{equation}
The effect of the spectator quark is calculated using the
Operator Product Expansion (OPE), which expands the
QCD corrections in powers of $\alphas(\mb)$ and
of $\LambdaQCD/\mb$.
The former, perturbative, corrections have been calculated to
$O(\alphas^2)$.\cite{vanRitbergen}
The latter, non-perturbative, corrections start at
$O(\LambdaQCD^2/\mb^2)$.
The leading term appears as
$(\lambdaone-9\lambdatwo)/(2\mb^2)$,
where $\lambdaone$ and $\lambdatwo$ are the parameters
due to the $b$ quark's Fermi motion and the hyperfine
interaction between the $b$ and light quarks, respectively.
The non-perturbative parameters $\lambdaone$ and $\lambdatwo$
are known from the
measurements of $\BtoXclnu$ decays and, in the case of
$\lambdatwo$, from the $B$-$B^*$ mass difference.
Overall, the uncertainties in $\Gamma(\BtoXulnu)$
due to the QCD corrections are small
compared with the uncertainty due to the
$b$ quark mass, which appears as $\mb^5$ in (\ref{eq:Gquark}).

The same OPE framework applied to the $\BtoXclnu$ decays
can predict $\Gamma(\BtoXclnu)$
and the moments of the kinematical variables
such as the lepton energy $\El$ and the hadronic mass $\mX$
in terms of $\Vcb$, $\alphas(\mb)$, $\mb$, $\mc$
and the non-perturbative parameters.
%Since the $\BtoXclnu$ decays are more precisely measured than
%the $\BtoXulnu$ decays,
%one must consider higher-order terms in the expansion,
%which introduce additional non-perturbative parameters.
By measuring the inclusive rate and several of the $\El$ and
$\mX$ moments as functions of the minimum lepton energy,
\babar\cite{BaBarOPEfit}
determined $\Vcb$, $\mb$, $\mc$, and the
non-perturbative parameters from a simultaneous fit to the
OPE calculation.
The results are
\begin{eqnarray*}
\Vcb  &=& (41.4\pm0.4\pm0.4\pm0.6)\times10^{-3},\\
\mb   &=& (4.61\pm0.05\pm0.04\pm0.02)\gev,\\
\mc   &=& (1.18\pm0.07\pm0.06\pm0.02)\gev,\\
\mupi &=& (0.45\pm0.04\pm0.04\pm0.01)\gev^2,\\
\muG  &=& (0.27\pm0.06\pm0.03\pm0.02)\gev^2,
\end{eqnarray*}
where the errors are experimental, uncertainties in the OPE
calculation, and other theoretical uncertainties, respectively.
The parameters $\mupi$ and $\muG$ are related to $\lambdaone$
and $\lambdatwo$, respectively.
The quark masses and the non-perturbative parameters
depend on the quark mass scheme and the renormalization scale
used in the OPE calculation;
the calculation\cite{GambinoUraltsev} adopted in this analysis
employed the kinetic scheme,\cite{BigiKinetic}
in which the non-perturbative contribution to the $b$-quark
mass is subtracted using heavy-quark sum rules,
with the scale $\mu=1.0\gev$.
The fit describes the data points quite well with $\chi^2=20$ for
15 degrees of freedom.
Bauer \etal\cite{BauerOPEfit} have performed a more extensive
global fit using measurements from
\babar, Belle, CLEO, CDF, and DELPHI\@.
Using the $1S$ scheme,\cite{1Sscheme}
which relates the $b$-quark mass to the mass
of the $\Upsilon(1S)$ resonance, they find
$\Vcb = (41.4\pm0.6\pm0.1)\times10^{-3}$,
where the first error includes both experimental and theoretical
uncertainties and the second error is due to the $B$ meson lifetime.
Bauer \etal\ repeat the fit with different expansion and mass schemes
and find good agreement.
In addition to determining \Vcb\ to a 2\% precision,
these fits demonstrate the reliability of the OPE framework in
predicting the inclusive decay rate and spectral moments.

Because of the presence of the $\BtoXclnu$
background, the inclusive $\BtoXulnu$ decay width
cannot be directly measured.
The experiments measure, instead, partial decay
widths in limited regions of the phase space that
are free from the $\BtoXclnu$ background.
This is achieved by a cut on one or more of the three
kinematic variables of the $X\lnu$ final state:
the lepton energy $\El$, the hadronic mass $\mX$, and
the lepton-neutrino invariant mass squared $q^2$.
The fraction, $\fu$, of the $\BtoXulnu$ events that
pass the experimental cut needs to be accurately
known in order to determine $\Vub$.

The OPE framework can reliably predict the inclusive
$B\to X_u\lnu$ decay rate as long as it is integrated
over a large region of the phase space.
The experimental cuts required to suppress the
$\BtoXclnu$ background violate this requirement.
In order to overcome this limitation,
a so-called twist expansion\cite{Twist}
is performed.
The leading term of the non-perturbative correction becomes
$O(\LambdaQCD/\mb)$ instead of $O(\LambdaQCD^2/\mb^2)$,
and is described by the distribution function, known
as the \emph{shape function}, of the light-cone momentum
of the $b$ quark inside the $B$ meson.

The shape function cannot be computed perturbatively,
and must be determined experimentally.
This is achieved by two methods:
\begin{itemlist}
\item
  The first and second moments of the shape function are
  related to $\Lambdabar = m_B - m_b$ and $\lambdaone$,
  which are determined by the OPE fit to the $B\to X_c\lnu$
  moments.
  The recent two-loop calculation by Neubert\cite{schemeNeubert}
  allows the translation of $m_b$ and $\mupi$ determined in various
  mass schemes into the shape-function scheme.
  Using the results from the \babar\ OPE fit discussed above,
  Neubert finds $m_b = (4.63\pm0.08)\gev$ and
  $\mupi = (0.15\pm0.07)\gev^2$ at the renormalization
  scale $\mu=1.5\gev$.
\item
  The photon energy spectrum in the $\btosg$ decays
  is affected, in the leading order of $\LambdaQCD/\mb$,
  by the same shape function.
  The $\btosg$ measurements by CLEO,\cite{CLEOellipse}
  Belle,\cite{Belleellipse} and \babar\cite{BABARbsg}
  are used to constrain $\Lambdabar$ and $\lambdaone$.
  The latest \babar\ measurement using the sum of
  exclusive $B\to X_s\gamma$ decays finds
  $m_b = (4.65\pm0.04)\gev$ and
  $\mupi = (0.19\pm0.06)\gev^2$
  in the shape-function scheme with $\mu=1.5\gev$.
\end{itemlist}
The good agreement between the results obtained by the
two independent methods suggests that the theoretical
uncertainties are under control.

In most of the $\BtoXulnu$ events that are experimentally
accessible, the hadronic system $\Xu$ has a small mass and
large momentum, i.e., it is jet-like.
The suitable theoretical tool, Soft Collinear Effective Theory
(SCET) has been developed in the last few years.\cite{SCET}
Recent theoretical analyses of the $\BtoXulnu$ and
$\btosg$ decays use SCET\@.
Until recently, the experiments relied on an $O(\alphas)$ OPE
calculation by De Fazio and Neubert\cite{DFN} to evaluate
the acceptance $\fu$ of their event selection criteria.
Latest results of \Vub\ extracted from inclusive measurements
use a new SCET-based calculation by Bosch \etal\cite{BLNP,LNP}
The new calculation tends to predict slightly larger $\fu$
than the previous calculation.
More importantly, the values of $\fu$ depend more strongly
on the shape-function parameters, making them the largest
source of uncertainty on \Vub\ as it will be seen
in Section~\ref{ssec:inclext}.

Beyond the shape-function uncertainty,
which we can expect to improve with better measurements
of $B\to X_c\lnu$ and $\btosg$,
a few theoretical issues remain unresolved:
\begin{itemlist}
\item
  The $B\to X_u\lnu$ and $\btosg$ processes are
  governed by the single shape function only in the leading
  order of $\LambdaQCD/m_b$, and the next-to-leading order
  corrections differ.
  The impact of the sub-leading shape functions
  is a subject of active theoretical
  research.\cite{SubSF}
\item
  For $\Bz\to X_u\lnu$,
  additional complication arises
  from the weak-annihilation diagram,\cite{BigiWA}
  in which the $b$ and $\dbar$ quarks annihilate into a $W^-$ boson.
  The contribution to the total rate is expected to be small
  ($\le$2\%), but concentrated near the lepton-energy endpoint
  where its relative contribution may be significant.
  The size of the effect of the weak annihilation can be
  constrained in the future
  by measuring the lepton spectrum separately for
  $\Bz\to\Xu\lnu$ and $\Bp\to\Xu\lnu$.
  Alternatively, a cut that rejects the highest-$q^2$ region,
  as proposed by Lange \etal,\cite{LNP}
  can be used to suppress the effect of the weak annihilation.
\end{itemlist}
The current estimates of these uncertainties contribute to
a theoretical error of $\sim$5\% on \Vub.
Understanding of these issues will become important
as we improve the experimental and shape-function errors
in the next few years.

\subsection{Experimental Measurements}\label{ssec:inclmeasure}

Recent inclusive measurements of \Vub\ use one of
the three techniques:
\begin{itemlist}
\item The lepton endpoint measurements\cite{BaBarEndPoint,BelleEndPoint}
  use the momentum spectrum of the leptons near the kinematical endpoint.
\item The neutrino reconstruction measurement\cite{BaBarQ2El}
  estimates the neutrino momentum from the missing momentum of the event.
\item The hadronic recoil measurements\cite{BelleVubtalk,BaBarVubBreco}
  use the recoil of the $B$ mesons that are fully reconstructed
  in hadronic decays.
\end{itemlist}

The upper endpoint region of the lepton-energy spectrum
offers the most accessible window
to the $B\to X_u\lnu$ signal.
The earliest measurements of the $\BtoXulnu$ decay by
CLEO\cite{CLEOobs} and
ARGUS\cite{ARGUSobs} 
used leptons with momenta beyond the kinematical
endpoint for the $\BtoXclnu$ decay.
The available phase space above the charm endpoint
is unfortunately small ($\fu\approx6\%$)
and extends beyond the quark-level endpoint.
The accessible signal fraction $\fu$ is therefore
strongly dependent on the shape function.
The signal rate near the endpoint is also sensitive
to the weak-annihilation effect.
These factors make it difficult to evaluate reliably
the theoretical uncertainty of $\fu$ for an endpoint measurement.

Recent measurements of $\Vub$ using the $\El$ spectrum
try to ameliorate the theoretical issues by extending the
signal region significantly below the $B\to X_c\lnu$ endpoint.
\babar\cite{BaBarEndPoint} and Belle\cite{BelleEndPoint}
measured the partial branching fraction $\Delta\BR(B\to X_ue\nu)$
for the electron momentum interval of $2.0$--$2.6\gev$
and $1.9$--$2.6\gev$, respectively.
The minimum lepton momentum in the previous measurements
from CLEO\cite{CLEOendpoint} and Belle\cite{Belleendpointold}
were $2.2\gev$ and $2.3\gev$, respectively.
The accessible signal fraction $\fu$ for $\El>2.0\gev$ is
as large as 28\%.

In order to achieve measurements far below the $\BtoXclnu$
threshold, accurate modeling and subtraction of the background,
both $\BtoXclnu$ and non-\BB, are essential.
The non-\BB\ background is subtracted using off-peak data
collected at center-of-mass energies below the $\Upsilon(4S)$
resonance.
The $E_e$ spectrum is then fitted with a combination of the
$\BtoXulnu$ signal, $B\to D\lnu$, $B\to\Dstar\lnu$,
$B\to D^{**}\lnu$, and non-resonant $B\to D^{(*)}\pi\lnu$
background distributions.
The convergence of the fit is helped by the fact
that the high end of the $E_e$ spectrum is dominated
by the $D$ and $\Dstar$ states.
Extending the $E_e$ ranges lower would require better
understanding of the production of the higher $D$ resonances
and non-resonant semileptonic decays.

Using data samples corresponding to $80\invfb$ and $27\invfb$,
\babar\ and Belle measure partial branching fractions
\begin{eqnarray*}
\Delta\BR(p_e>2.0\gev)
  &=& (5.31\pm0.32\pm0.49)\times10^{-4},\\
\Delta\BR(p_e>1.9\gev)
  &=& (8.47\pm0.37\pm1.53)\times10^{-4},
\end{eqnarray*}
respectively, where the errors are statistical and systematic.
The larger systematic error of the Belle result reflects the
difficulty associated with understanding the increasing
background at lower lepton energies.

The neutrino-reconstruction measurement by \babar\cite{BaBarQ2El}
combines electrons with $E_e>1.9\gev$ with the neutrino momentum,
inferred from the missing momentum of the event,
to calculate the $q^2$ of the lepton-neutrino system.
For a given set of $(E_e,q^2)$, the maximum hadronic mass
squared can be calculated as
\begin{equation}
\shmax = m_B^2 + q^2 - 2m_B\left(E_e + \frac{q^2}{4E_e}\right),
\end{equation}
plus a small correction to account for the $B$ momentum in
the $\Upsilon(4S)$ rest frame.\cite{KowaMenke}
Requiring $\shmax<3.5\gev^2$ removes a large fraction of the
$\BtoXclnu$ background.
\babar\ achieves a signal-to-background ratio of $1/2$
with a signal acceptance $f_u\sim16\%$.
The resolution of the $\shmax$ variable is studied using
a control sample of exclusively reconstructed $B\to D^{(*)}\lnu$ decays.
The measured partial branching fraction is
\[
\Delta\BR(E_e>1.9\gev,\shmax<3.5\gev^2) = (4.46\pm0.42\pm0.83)\times10^{-4},
\]
where the errors are statistical and systematic.
The data sample used corresponds to $80\invfb$.
The systematic error is expected to improve by the time the
measurement is published.

The large signal acceptances of the $\mX$ and $q^2$ cuts
alleviate some of the theoretical difficulties encountered
by the $\El$ endpoint measurements.
In principle, an $\mX$ cut at the $D$ meson mass provides
the best signal acceptance possible ($\fu\approx70\%$).
The $\mX$ spectrum, however, has a strong shape-function
dependence near the $D$ threshold, and so does $\fu$.
A $q^2$ cut at $(m_B-m_D)^2$, on the other hand,
provides a moderate efficiency ($\fu\approx20\%$) with
small dependence on the shape function.
Two other cuts have been proposed to minimize the theoretical
errors while maintaining good signal acceptance:
\begin{itemlist}
\item
  Bauer \etal\cite{BauermXq2cut} proposed
  a combined cut on both $\mX$ and $q^2$
  that would maximize $\fu$
  while minimizing its shape-function dependence.
  For a typical set of cuts at $\mX<1.7\gev$ and
  $q^2>8\gev$, the acceptance $\fu$ is around 30\%
  with an uncertainty of $\sim6\%$.
\item
  Mannel and Recksiegel,\cite{MannelP+}
  and more recently
  Bosch \etal,\cite{BLNP,BLNP2} proposed a cut on
  $\Pplus = E_X - p_X$, where $E_X$ and $p_X$ are the
  energy and momentum of the hadronic system, respectively.
  A cut at $\Pplus<m_D^2/m_B$ would have a large
  $\fu$ of $\sim$60\% and reject the region where the
  SCET is not applicable.
\end{itemlist}

In order to measure the hadronic mass $\mX$,
one must identify all the decay products of the $B$ meson.
This is achieved in the hadronic-recoil measurements by
\babar\cite{BaBarVubBreco} and Belle\cite{BelleVubtalk}
by completely reconstructing one $B$ meson
and using the recoiling $B$ meson.
With full reconstruction of hadronic $B$ decays,
the four-momentum of the recoil $B$ is known by momentum
conservation.
After finding a lepton candidate with $p_{\ell}>1\gev$
in the recoil,
a kinematical fit calculates the most likely four-momenta
for the neutrino and for the hadronic system.
The typical $\mX$ resolution is $350\mev$.
The $\BtoXclnu$ background is suppressed by the likely
presence of kaons in the final state, and by finding
soft pions that are kinematically consistent with the
$B\to\Dstar\lnu$ decay.

The hadronic-recoil measurements can measure the partial
branching ratio in a variety of regions of the $(\mX,\qq)$
space.
For $\mX<1.7\gev$ and $q^2>8\gev^2$, \babar\ and Belle find
\begin{eqnarray*}
\Delta\BR(\mX<1.7\gev,\qq>8\gev^2)
  &=& (8.96\pm1.43\pm1.44)\times10^{-4},\\
\Delta\BR(\mX<1.7\gev,\qq>8\gev^2)
  &=& (8.41\pm0.85\pm1.03)\times10^{-4},
\end{eqnarray*}
respectively,
where the errors are statistical and systematic.
The data samples used are $80\invfb$ for \babar, and
$253\invfb$ for Belle.
Belle has also measured the partial branching fraction
\[
\Delta\BR(\Pplus<0.66\gev) = (11.0\pm1.0\pm1.6)\times10^{-4},
\]
where the errors are statistical and systematic.

\subsection{Extraction of \Vub}\label{ssec:inclext}

The partial branching fractions measured by the experiments
can be translated into \Vub\ by
\begin{equation}
\Vub = \sqrt{\frac{\Delta\BR}{\Gthy\tau_B}}, \label{eq:extract}
\end{equation}
where $\tau_B$ is the average lifetime of $\Bz$ and $\Bp$.
The reduced decay rate $\Gthy$ is defined as
\begin{equation}
\Gthy \equiv \frac{\Delta\Gamma_{\mathrm{thy}}}{\Vub^2},
\end{equation}
where $\Delta\Gamma_{\mathrm{thy}}$ is the partial width
of the $\BtoXulnu$ decay into the phase space of interest
predicted by the theory.

Until recently, the conversion was often performed in two steps.
The measured $\Delta\BR$ was converted into the total 
branching fraction
$\BR = \Delta\BR/\fu$,
which was then converted into $\Vub$ using the OPE calculation
for the fully inclusive decay rate $\Gamma(\BtoXulnu)$.
Since both $\fu$ and $\Gamma(\BtoXulnu)$ depend critically
on $m_b$, care must be taken to treat the correlation in their
uncertainties.
Equation (\ref{eq:extract}) avoids this problem by
calculating the partial decay rate directly.\cite{LNP}

\begin{figure}
  \begin{center}
    \psfig{file=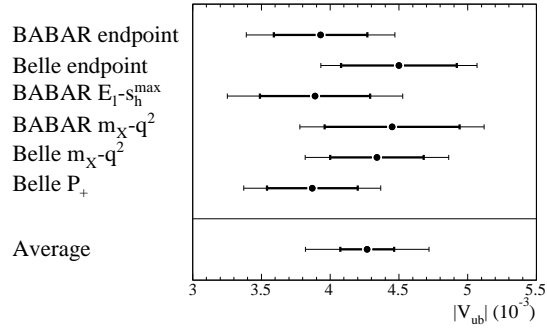,width=0.6\textwidth}
  \end{center}
  \caption{\label{fig:inclvub}
    Results of the inclusive \Vub\ measurements.
    The thick parts of the error bars indicate the experimental
    errors.
    The average excludes the Belle $\Pplus$ result.}
\end{figure}

Table~\ref{tab:inclvub} and Fig.~\ref{fig:inclvub}
summarizes the values of \Vub\ extracted
by \babar\cite{BaBarVubtalk} and Belle\cite{BelleVubtalk}
from the inclusive measurements.
\begin{table}
  \tbl{\label{tab:inclvub}
    Measured partial branching fractions $\Delta\BR$
    and the extracted values of \Vub.
    The errors on $\Delta\BR$ are statistical and systematic.
    The errors on $\Vub$ are experimental, due to the shape function, and
    theoretical.
    The average excludes the Belle $\Pplus$ result.}
  {\begin{tabular}{@{}lcrc@{}}
    \toprule
    & Cuts & \multicolumn{1}{c}{$\Delta\BR\times10^4$} & $\Vub\times10^3$ \\
    \colrule
    \babar\ & $\El>2.0\gev$                   & $5.3\pm0.3\pm0.5$ & $3.93\pm0.34\pm0.38\pm0.18$ \\ 
    Belle   & $\El>1.9\gev$                   & $8.5\pm0.4\pm1.5$ & $4.50\pm0.42\pm0.32\pm0.21$ \\
    \babar\ & $\El>1.9\gev, \shmax<3.5\gev^2$ & $4.5\pm0.4\pm0.8$ & $3.89\pm0.40\pm0.45\pm0.21$ \\
    \babar\ & $\mX<1.7\gev, \qq>8\gev^2$      & $9.0\pm1.4\pm1.4$ & $4.45\pm0.49\pm0.40\pm0.22$ \\
    Belle   & $\mX<1.7\gev, \qq>8\gev^2$      & $8.4\pm0.9\pm1.0$ & $4.34\pm0.34\pm0.33\pm0.22$ \\
    Belle   & $\Pplus<0.66\gev$               & $11.0\pm1.0\pm1.6$ & $3.87\pm0.33\pm0.35\pm0.13$ \\
    \colrule
    \multicolumn{2}{@{}l}{Average \Vub} & & $4.27\pm0.20\pm0.35\pm0.21$ \\
    \botrule
  \end{tabular}}
\end{table}
Both experiments used the calculation by Bosch \etal\cite{BLNP,LNP}
to evaluate $\Gthy$.
For the shape function parameters, \babar\ used
$\mb=(4.63\pm0.08)\gev$ and $\mupi=(0.15\pm0.07)\gev^2$
derived from the $\BtoXclnu$ OPE fit,\cite{schemeNeubert}
and Belle used $\mb=(4.63\pm0.07)\gev$ and $\mupi=(0.20\pm0.07)\gev^2$.
The theoretical errors include perturbative errors, effects of the
subleading shape functions and weak annihilation.

The existing measurements agree well with each other.
Assuming that the experimental errors are uncorrelated and
the shape function and theory errors are fully correlated,
we find the average value
\[
\Vub = (4.27\pm0.20\pm0.35\pm0.21)\times10^{-3},
\]
where the errors are experimental, due to the shape function, and
theoretical, respectively.
The Belle $\Pplus$ result is not included in this average because of
the strong correlation with the Belle $\mX$-$\qq$ result.
The overall precision is $\pm10.5\%$.

The new central value is significantly smaller than
the 2004 summer average by the Heavy Flavor Averaging Group,
$\Vub=(4.70\pm0.44)\times10^{-3}$.
The change is mainly due to the shape-function
parameters, for which the old average used the values extracted
from the Belle $b\to s\gamma$ measurement.\cite{Belleellipse}
The uncertainty has grown
despite the availability of the more precise shape-function parameters.
This is partly due to the stronger shape-function dependence of the
new theoretical calculation, and also to the more careful assessment
of the theoretical uncertainties.\cite{LNP}

The precision of \Vub\ from the inclusive measurements is
limited by the uncertainties in the shape-function parameters,
in particular $m_b$.
The results quoted above assume $\sigma(m_b)=70$--$80\mev$,
which is probably conservative; as discussed earlier,
the latest measurements of $\btosg$ can determine $m_b$ to $\pm40\mev$.\cite{BABARbsg}
Using the data samples collected in the next few years,
\Vub\ may be determined with a precision of $\sim$7\%.

\section{Exclusive Measurements of \Vub}\label{sec:excl}

Exclusive measurements of \Vub\ can be performed with
$B\to\pi\lnu$, $\rho\lnu$, $\omega\lnu$, $\eta\lnu$, etc.
Among the possible channels,
the $\pi\lnu$ decay offers the cleanest path to \Vub\ both
experimentally and theoretically.

\subsection{Theoretical Background}\label{ssec:excltheory}

The differential decay rate of the $B\to\pi\lnu$ decay is
given by
\begin{equation}
  \frac{d\Gamma(B\to\pi\lnu)}{dq^2} =
  \frac{G_F^{2}\Vub^2}{24\pi^3}|f_+(q^2)|^2p_{\pi}^3,
\end{equation}
where $f_+(q^2)$ is the form factor.
We assume from isospin symmetry
\begin{equation}
  \Gamma(\Btopilnu) \equiv
  \Gamma(\Bztopilnu) = 2\Gamma(\Bptopilnu).
\end{equation}
The form factor $f_+(q^2)$ is significantly harder to calculate 
than the $B\to\Dstar$ and $B\to D$ form factors.
The estimates in literature employ a variety of techniques including
quenched lattice QCD (LQCD),
light-cone sum rules (LCSR),
quark model, and skewed parton distributions.
None of the techniques provide robust means of estimating the
associated uncertainties.

Two developments in 2004 considerably improved the situation.
\begin{itemlist}
\item Ball and Zwicky published an improved LCSR
  calculation.\cite{Ball04}
  They quote a total uncertainty of about 13\% 
  in the small-$q^2$ region ($q^2<14\gev^2$).
\item Two preliminary results of unquenched LQCD calculations
  were presented by the HPQCD\cite{HPQCD04}
  and FNAL\cite{FNAL04} collaborations.
  The quoted uncertainties at the large-$q^2$ region
  ($q^2>15\gev^2$) are about 13\% in both cases.
\end{itemlist}
The new calculations should allow reliable determination of
$\Vub$ from $\Gamma(\Btopilnu)$,
and stimulated renewed interest in such measurements.

Note that the LQCD and LCSR calculations are valid in limited
and non-overlap\-ping regions of $q^2$, namely
above $15\gev^2$ and below $14\gev^2$, respectively.
Although the authors provide the extrapolation of their
calculations to the full $q^2$ range, such extrapolations
add uncertainties that are not fully quantifiable.
It is therefore important for the experiments
to measure the differential decay rate
as a function of $q^2$.

\subsection{Experimental Measurements}\label{ssec:exclmeasure}

Four measurements of the differential decay rate
$d\Gamma(\Btopilnu)/dq^2$ have been reported by
CLEO,\cite{CLEOpilnu} Belle,\cite{Bellepilnu}
and recently by \babar\cite{BABARpilnu} using two
different techniques:
\begin{itemlist}
\item
  The neutrino reconstruction measurements
  reconstruct the neutrino as the missing four-momentum
  of the whole event.
\item
  The semileptonic recoil measurements
  use the recoil of the $B$ mesons
  tagged by their semileptonic decays.
\end{itemlist}
Another preliminary measurement by \babar\cite{BaBarVubBreco}
uses the same hadronic recoil technique used in the inclusive
\Vub\ measurement, and finds
$\BR(\Bztopilnu)=(1.08\pm0.28\pm0.16)\times10^{-4}$.
The statistics of this measurement is too small to allow binning
in $q^2$.

The neutrino-reconstruction technique has been
used by CLEO\cite{CLEOpilnu} and \babar.\cite{BABARpilnu}
The momentum of the neutrino is inferred from the missing
momentum in the event, and is combined with a lepton and a pion to form
a $\Btopilnu$ candidate.
The signal is observed as a peak in the $(m_B,E_B)$ plane.
The CLEO detector, with a better geometrical coverage
than the detectors at the $B$ Factories, is particularly suited
for this technique.
The main sources of background are misreconstructed
$\BtoXclnu$ events and cross-feed from the $B\to\rho\lnu$ decay.
The latter makes this type of measurements
susceptible to uncertainties of the $B\to\rho$ form factors.

The semileptonic-recoil technique has been used
by Belle\cite{Bellepilnu} and \babar.\cite{BABARpilnu}
The technique is similar to the hadronic-recoil technique,
but reconstructs the tag-side $B$ meson in the $B\to D\lnu$ or
$\Dstar\lnu$ decays.
The efficiency is higher than the hadronic reconstruction by
three to four times, thanks to the larger branching fractions
of the accessible decay channels.
On the other hand, the presence of the extra neutrino makes
the kinematics less strongly constrained, resulting in higher
background.
Overall, this technique provides an
efficiency and a signal-to-background ratio that are
well matched with the low branching fractions and moderate
background levels of the exclusive measurements.

\begin{figure}
  \begin{center}
    \psfig{file=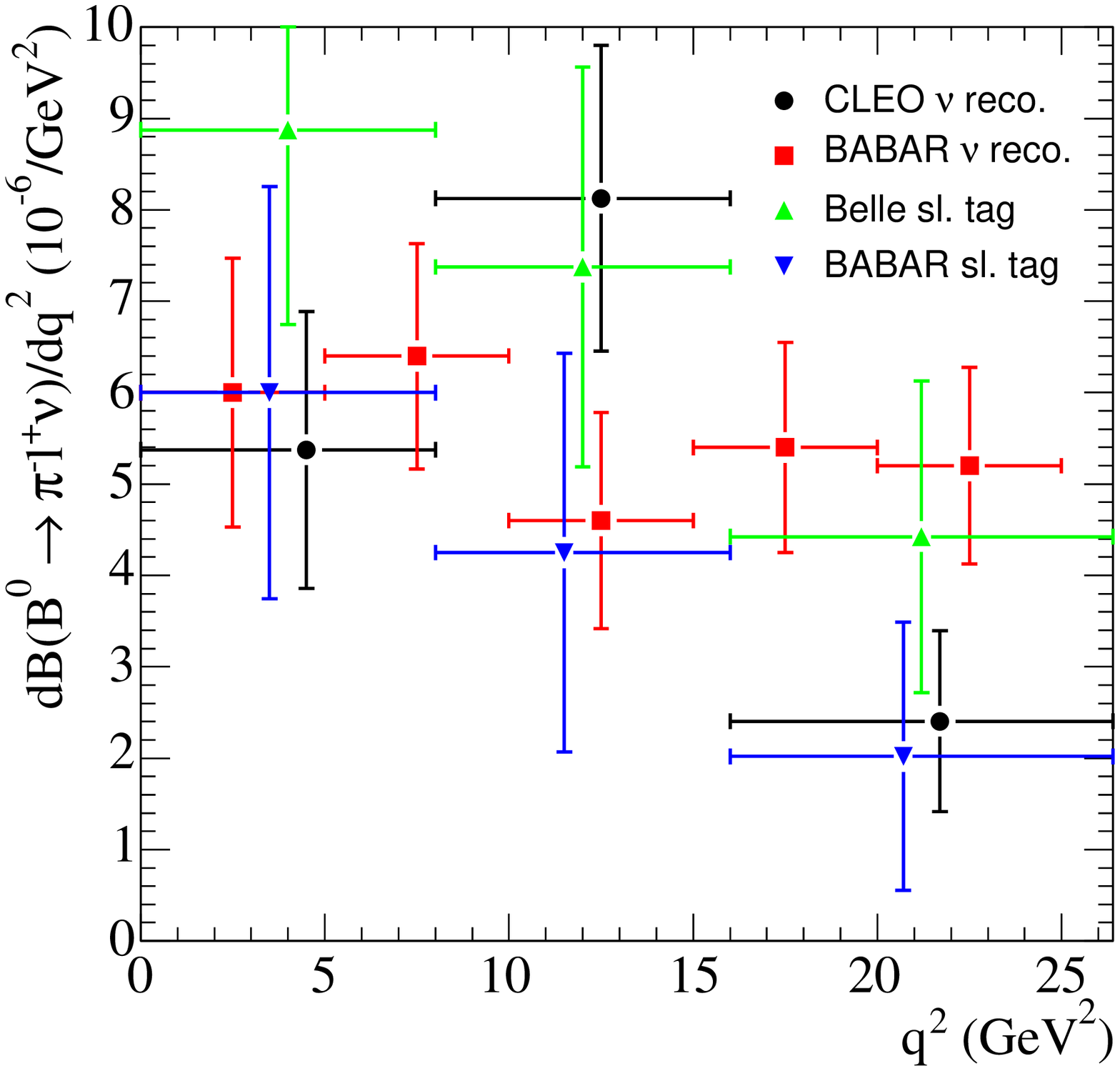,width=0.6\textwidth}
  \end{center}
  \caption[]{Measured $d\BR(\Bztopilnu)/dq^2$.}
  \label{fig:pilnu}
\end{figure}

The results of the four measurements agree reasonably well,
as shown in Fig.~\ref{fig:pilnu}.
Table~\ref{tab:pilnu} summarizes the partial branching
fractions in the low- and high-$q^2$ regions where
the LCSR and LQCD calculations are valid, respectively.
\begin{table}
  \tbl{\label{tab:pilnu}
    Measured $\Delta\BR(\Bztopilnu)$ in low and high
    regions of $q^2$.
    The errors are statistical, experimental
    systematic, and due to the $B\to\rho\lnu$ form factors.}
  {\begin{tabular}{@{}lccc@{}}
    \toprule
    & $q^2_{\mathrm{cut}}$
    & $\Delta\BR(q^2<q^2_{\mathrm{cut}})\times10^4$
    & $\Delta\BR(q^2>q^2_{\mathrm{cut}})\times10^4$ \\
    \colrule
    CLEO $\nu$ reco.   & $16\gev^2$
    & $1.08\pm0.14\pm0.09\pm0.04$ & $0.25\pm0.09\pm0.05\pm0.03$ \\
    \babar\ $\nu$ reco.  & $15\gev^2$
    & $0.85\pm0.08\pm0.11\pm0.05$ & $0.53\pm0.05\pm0.06\pm0.03$ \\
    Belle sl. tag & $16\gev^2$
    & $1.30\pm0.22\pm0.15\pm0.02$ & $0.46\pm0.17\pm0.05\pm0.01$ \\
    \babar\ sl. tag & $16\gev^2$
    & $0.82\pm0.23\pm0.12\pm0.00$ & $0.21\pm0.14\pm0.06\pm0.01$ \\
    \botrule
  \end{tabular}}
\end{table}
Note that the threshold separating the two $q^2$
regions is different for the \babar\ neutrino-reconstruction
measurement, which uses five $q^2$ bins instead of three.

\subsection{Extraction of \Vub}

\begin{figure}
  \begin{center}
    \psfig{file=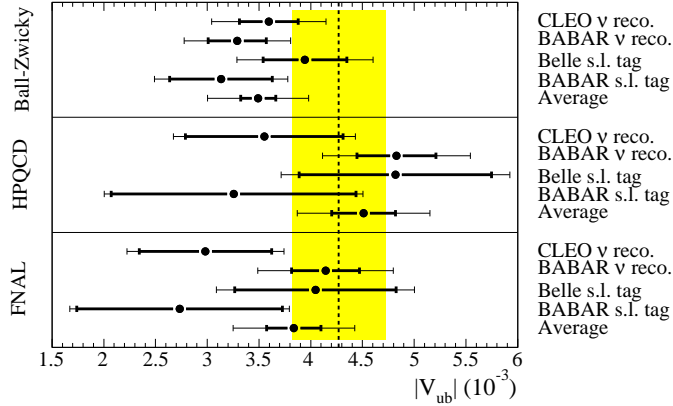,width=0.75\textwidth}
  \end{center}
  \caption{\label{fig:vubpilnu}
    Values of \Vub\ from the measurements of $\Delta\BR(\Bztopilnu)$.
    The thick part of the error bars correspond to the errors in
    the $\Delta\BR$ measurements.
    The dashed line and the shaded region indicate the average
    of the inclusive measurements discussed in Section~\ref{sec:incl}.}
\end{figure}

The measured branching fractions $\Delta\BR(\Bztopilnu)$
can be converted into \Vub\ using
Equation (\ref{eq:extract}) with $\tau_B$ replaced by the $\Bz$
lifetime.
The reduced decay rate $\Gthy$ is given by
\begin{equation}
\Gthy = \frac{G_F^2}{24\pi^3}
\int_{q^2_{\min}}^{q^2_{\max}} |f_+(q^2)|^2 p_{\pi}^3 dq^2.
\end{equation}
The values of $\Gthy$ calculated from Ball-Zwicky,\cite{Ball04}
HPQCD,\cite{HPQCD04} and FNAL\cite{FNAL04} are given in
Table~\ref{tab:gthy}.
\begin{table}
  \tbl{\label{tab:gthy}
    Values of $\Gthy$ from recent calculations of the
    $B\to\pi\lnu$ form factor.}
  {\begin{tabular}{@{}lcccc@{}}
    \toprule
    & $q^2$ ($\gev^2$) & $\Gthy$ ($\ps^{-1}$)
    & $q^2$ ($\gev^2$) & $\Gthy$ ($\ps^{-1}$) \\
    \colrule
    Ball-Zwicky
      & $<15$ & $5.11\pm1.34$ & $<16$ & $5.44\pm1.43$ \\
    HPQCD
      & $>15$ & $1.48\pm0.37$ & $>16$ & $1.29\pm0.32$ \\
    FNAL
      & $>15$ & $2.01\pm0.55$ & $>16$ & $1.83\pm0.50$ \\
    \botrule
  \end{tabular}}
\end{table}
The errors in $\Gthy$ reflects the form-factor uncertainties
quoted by the authors.
Applying the $\Gthy$ values to the measurements in Table~\ref{tab:pilnu},
we find the \Vub\ values in Table~\ref{tab:vubpilnu} and Fig.~\ref{fig:vubpilnu}.
\begin{table}
  \tbl{\label{tab:vubpilnu}
    Values of \Vub\ in $10^{-3}$
    from the measurements of $\Delta\BR(\Bztopilnu)$
    and the recent calculations of the form factor.
    The errors are due to the $\Delta\BR$ measurements and to the
    form factor calculations.}
  {\begin{tabular}{@{}lccc@{}}
    \toprule
    & Ball-Zwicky & HPQCD & FNAL \\
    \colrule
    CLEO $\nu$ reco.
      & $3.60\pm0.28\pm0.47$ & $3.55\pm0.76\pm0.44$ & $2.98\pm0.64\pm0.41$ \\
    \babar\ $\nu$ reco.
      & $3.29\pm0.28\pm0.43$ & $4.83\pm0.38\pm0.60$ & $4.14\pm0.33\pm0.57$ \\
    Belle sl. tag
      & $3.94\pm0.41\pm0.52$ & $4.82\pm0.93\pm0.60$ & $4.05\pm0.78\pm0.55$ \\
    \babar\ sl. tag
      & $3.13\pm0.50\pm0.41$ & $3.26\pm1.18\pm0.40$ & $2.73\pm0.99\pm0.37$ \\
    \colrule
    Average
      & $3.49\pm0.17\pm0.46$ & $4.51\pm0.31\pm0.56$ & $3.84\pm0.26\pm0.53$ \\
    \botrule
  \end{tabular}}
\end{table}
The first errors come from the partial branching fraction measurements,
and the second from the form-factor calculations.
The average \Vub\ values in Table~\ref{tab:vubpilnu} were calculated
assuming no correlations between the experimental errors.
Even for the two \babar\ measurements, the overlap between the
signal samples is small.

Considering only the experimental errors,
the measurements of $B\to\pi\lnu$ partial branching
fractions can determine \Vub\ with a precision better than $\pm7\%$.
The agreement among the existing measurements is satisfactory.
The differences among the available form-factor calculations
remain considerable,
although they are consistent with the quoted theoretical uncertainties.

Historically,
the values of \Vub\ extracted from the exclusive measurements
tended to be smaller than the results of the inclusive measurements.
This is no longer the case with the latest measurements,
partly because the inclusive results have gone down,
and partly because the new LQCD results are higher than the
older exclusive measurements.

Further improvements in the form-factor calculation
are necessary to bring the uncertainty on \Vub\ below
10\% level.
The leading sources of the uncertainties are
operator matching and finite lattice spacing for the HPQCD and
FNAL calculations, respectively.
Technical improvements to overcome these limitations may
reduce the total theoretical errors to 5--6\% level
in the future.\cite{ShigemitsuCKM2005}

\subsection{Other Exclusive Measurements}\label{ssec:exclother}
 
In addition to $B\to\pi\lnu$, the experiments have measured
the decay rates for
$B\to\rho\lnu$,\cite{CLEOpilnu,Bellepilnu,BABARpilnu,CLEOrholnu,BABARrholnu}
$B\to\eta\lnu$,\cite{CLEOpilnu} and
$B\to\omega\lnu$.\cite{Belleomegalnu}
While the $\rho\lnu$ mode has a larger rate than the
$\pi\lnu$ mode, one must deal with the non-resonant
$\pi\pi$ contribution.
Theoretically, the $\rho\lnu$ mode suffers from the lack
of techniques that take into account the width of the $\rho$
resonance.
Much progress in LQCD will be necessary before
the $B\to\rho\lnu$ decay can be used to extract $\Vub$ reliably.
The $\eta\lnu$ and $\omega\lnu$ channels,
while more challenging to measure,
are expected to be more tractable from the theoretical
point of view, and may provide valuable cross-checks in
the future.

\section{Summary and Outlook}\label{sec:sum}

A precise determination of \Vub\ is one of the most sought-after
physics goals pursued at the $B$ factories.
The field of charmless semileptonic $B$ decays have seen
rapid experimental and theoretical progresses in the last few years.
Innovative experimental techniques combined with new theoretical
insights have brought the relative precision of \Vub\ close to $\pm10\%$.

Through the inclusive measurements, we find an average
value of
$\Vub=(4.27\pm0.45)\times10^{-3}$.
The error is dominated by the uncertainty in the shape function,
which can be improved by measurements of the $\BtoXclnu$ spectra
and of the $\btosg$ spectrum.
The exclusive $\Btopilnu$ measurements give consistent values
of \Vub.
The errors are dominated by the $\pm13\%$ theoretical
uncertainties in the form factor.

A 10\% precision in \Vub\ is almost certaintly around the corner.
In a few years, with half a billion $\BB$ events per experiment, 
it is likely that we will be able to determine \Vub\ to a precision
of $\sim$7\% through the inclusive measurements.
Progress in LQCD will eventually achieve a similar
precision through the exclusive $\Btopilnu$ measurements,
whose experimental precision on \Vub\ is already at a 7\% level.

\section*{Acknowledgments}

It is my pleasure to thank P.~Fisher, R.~Kowalewski and G.~Sciolla
for their kind help in preparing this article.
This work was supported in part by the DOE contract
DE-FG02-91ER40654.

%%% citation symbols %%%
\newcommand{\PRL}[3]{Phys.\ Rev.\ Lett.\ \textbf{#1}, #2 (#3)}
\newcommand{\PRD}[3]{Phys.\ Rev.\ D \textbf{#1}, #2 (#3)}
\newcommand{\PLB}[3]{Phys.\ Lett.\ B \textbf{#1}, #2 (#3)}
\newcommand{\EPJC}[3]{Eur.\ Phys.\ J. C \textbf{#1}, #2 (#3)}
\newcommand{\PTP}[3]{Prog.\ Theor.\ Phys.\ \textbf{#1}, #2 (#3)}
\newcommand{\IJMP}[3]{Int.\ J.\ Mod.\ Phys.\ \textbf{#1}, #2 (#3)}
\newcommand{\JHEP}[3]{JHEP \textbf{#1}, #2 (#3)}
\newcommand{\NPB}[3]{Nucl.\ Phys.\ B \textbf{#1}, #2 (#3)}
\newcommand{\AIPCP}[3]{AIP Conf.\ Proc.\ \textbf{#1}, #2 (#3)}
\newcommand{\NPPS}[3]{Nucl.\ Phys.\ Proc.\ Suppl.\ \textbf{#1}, #2, (#3)}
\newcommand{\ARGUSColl}{ARGUS Collaboration}
\newcommand{\BABARColl}{\babar\ Collaboration}
\newcommand{\BelleColl}{Belle Collaboration}
\newcommand{\CLEOColl}{CLEO Collaboration}
\newcommand{\HFAG}{Heavy Flavor Averaging Group}

\end{document}